\documentclass[%
 reprint,
 amsmath,amssymb,
 aps,
 prr,
 floatfix,
 footnoteinthebib,
 longbibliography
]{revtex4-2}

\usepackage{graphicx,physics}
\usepackage{color}
\usepackage{dcolumn}
\usepackage{latexsym}

\usepackage[normalem]{ulem}
\usepackage{hyperref,amssymb}
\usepackage{url}
\usepackage{color,soul}
\usepackage{graphicx}
\usepackage{verbatim}
\usepackage{multirow}
\usepackage{amsmath}
\newcommand{\beq}{\begin{eqnarray}}
\newcommand{\eeq}{\end{eqnarray}}
\usepackage{mathrsfs}
\usepackage{soul}
\usepackage[dvipsnames]{xcolor}
\usepackage{mathtools}
\usepackage{slashed}
\usepackage{physics}	
\usepackage{graphicx}   
\usepackage{epstopdf}
\usepackage{subfigure}  
\usepackage{hyperref}   
\usepackage{bbold}
\usepackage{wasysym}
\usepackage{feynmp}
\usepackage{hyperref}
\usepackage{float}
\hypersetup{colorlinks,
}
\usepackage{bm}
\DeclareMathOperator{\sign}{sign}
\newcommand{\R}{\ensuremath{\mathbb{R}}}
\newcommand{\I}{\ensuremath{\mathbb{I}}}

\usepackage[latin1]{inputenc}

\begin{document}

\title{Non-Hermitian wave-packet dynamics and its realization \\within a non-Hermitian chiral cavity}
\author{Weicen Dong$^{1,2,3}$}
\email{orange233@sjtu.edu.cn}
\author{Qing-Dong Jiang$^{1,4,5}$}
\email{qingdong.jiang@sjtu.edu.cn}
\author{Matteo Baggioli$^{1,2,3}$}
\email{b.matteo@sjtu.edu.cn}
\address{$^1$School of Physics and Astronomy, Shanghai Jiao Tong University, Shanghai 200240, China}
\address{$^2$Wilczek Quantum Center, School of Physics and Astronomy, Shanghai Jiao Tong University, Shanghai 200240, China}
\address{$^3$Shanghai Research Center for Quantum Sciences, Shanghai 201315,China}
\address{$^4$Tsung-Dao Lee Institute, Shanghai Jiao Tong University, Shanghai 200240, China}
\address{$^5$Shanghai Branch, Hefei National Laboratory, Shanghai 201315, China}

\begin{abstract}

Topological wave-packet dynamics provide a powerful framework for studying quantum transport in topological materials. However, extending this approach to non-Hermitian quantum systems presents several important challenges, primarily due to ambiguities in defining the Berry phase and the non-unitary evolution of the wave-packets when $\mathcal{P}\mathcal{T}$ symmetry is broken. In this work, we adopt the complex Berry phase definition using the bi-orthogonal formalism and derive the semiclassical equations of motion (EOM) for a wave-packet in a non-Hermitian topological system. Interestingly, we find that the complex Berry curvature introduces both an anomalous velocity and an anomalous force into the semiclassical EOM. To validate the derived EOM, we design a non-Hermitian Haldane model featuring non-reciprocal next-nearest-neighbor (NNN) hopping, where the imbalance in the NNN hopping amplitudes gives rise to an emergent `\textit{complex chirality}'. We reveal that the real and imaginary components of the complex chirality dictate the signs of both the real and imaginary parts of the complex Berry curvature, as well as the direction and dissipation rate of the edge states. Our analytical findings are confirmed by direct numerical simulations of the wave-packet dynamics. Finally, we suggest a potential experimental realization of this complex Haldane model using a non-Hermitian optical chiral cavity, providing a promising platform for testing our theoretical predictions.
\end{abstract}

 \maketitle

\section{Introduction}

The concept of topology has been integral to the development of physics, tracing back to the early advancements in electromagnetism \cite{nash1999topology}. In recent decades, the discovery of measurable effects induced by topology, \textit{e.g.} the quantum Hall effect \cite{RevModPhys.82.3045, RevModPhys.83.1057}, has led to the blossom of condensed matter physics. Unlike non-topological systems, physical degrees of freedom (electrons, for example) in topological materials experience finite Berry curvature, which alters their semiclassical equations of motion (EOM). Electrons in solids are best described as wave packets with well-defined central momentum and position, superposed by Bloch wave functions. This semiclassical approach not only provides a practical tool for calculating transport properties but also offers a powerful framework for further understanding and quantizing physical systems \cite{PhysRev.95.1154, PhysRevLett.75.1348, PhysRevB.53.7010, Girvin_Yang_2019,Chong2010}.

In recent years, the long-held belief that a physical Hamiltonian must be Hermitian has been challenged \cite{Bender_2007}. By relaxing the Hermicity constraint and combining it with topology, many fascinating phenomena that are absent in Hermitian systems, such as non-Hermitian skin effects, have emerged  \cite{PhysRevResearch.6.043023, PhysRevResearch.4.023160,  FengLiang, okuma2023non,Ashida02072020, RevModPhys.93.015005, PhysRevX.9.041015, Ghatak_2019, PhysRevX.8.031079,PhysRevLett.123.206404,  PhysRevLett.118.040401, PhysRevLett.121.136802,MartinezAlvarez2018, Ding2022, PhysRevLett.123.066404, PhysRevLett.121.086803, PhysRevLett.115.040402, PhysRevLett.124.056802, PhysRevB.84.205128, PhysRevLett.125.126402}.
However, despite the rapid advancements in non-Hermitian topological systems, a semiclassical EOM for carriers---which is crucial for understanding Hermitian topological systems---remains an open question in this context.

In this work, we derive the semiclassical EOM for non-Hermitian topological systems with finite and complex Berry curvature. A non-Hermitian Hamiltonian features distinct left- and right-eigenvectors  (i.e., $|\psi_L(\textbf{k})\rangle$ and $|\psi_R(\textbf{k})\rangle$) that are independent of each other. To define the Berry connection, we employ the bi-orthogonal formalism where the Berry connection is given by $\bold A(\bold k)=\mathrm{i}\langle \psi_L(\textbf{k})|\nabla_\bold k|\psi_R(\textbf{k})\rangle$ \cite{Sun1993}. Such an approach offers the advantage of yielding time-independent Berry curvature, contrasting with previous definitions that rely on either left or right eigenvectors \cite{Hu:24, hu2024rolequantumgeometrictensor, PhysRevB.102.245147, PhysRevResearch.6.023202, wang2022anomalous}. 

To validate the derived semiclassical EOM,
we investigate wave-packet dynamics in a non-Hermitian Haldane model with non-reciprocal next-nearest-neighbor (NNN) hoppings \cite{PhysRevResearch.2.013387, wu, PhysRevB.84.205128} and further explore the novel relationships between complex Berry curvature and topological edge states. 
Our results reveal that both imaginary energy (as shown in classical dissipative systems \cite{muschietti1993}) and imaginary Berry curvature influence the EOM, leading to potential experimental observable transport signatures (see Eqs. \eqref{eq:eom_bulk_r} and \eqref{eq:eom_bulk_k}).  Additionally, we employ numerical simulations as a complementary approach to study the non-unitary evolution of non-Hermitian topological wave packets.
It is important to note that the non-Hermitian Haldane model used in this work differs from previous studies on non-Hermitian topological insulators \cite{okuma2023non, Ashida02072020, RevModPhys.93.015005, Ghatak_2019, PhysRevX.9.041015, PhysRevX.8.031079, PhysRevLett.121.136802, PhysRevLett.123.206404, PhysRevLett.118.040401, PhysRevLett.121.136802} and non-Hermitian Haldane models with complex on-site energy \cite{PhysRevResearch.2.013387, wu, PhysRevB.84.205128}. This model features ``complex chirality" and could potentially be realized through strong light-matter interactions in non-Hermitian photonic chiral cavities \cite{Bylinkin2021, Itai2020, Song2005,jiang2023, yang2024}.

\section{Non-Hermitian wave-packet dynamics}
In this section, we derive the EOM for a wave-packet in a non-Hermitian system which exhibits complex Berry curvature. The derivation relies on the non-unitary time evolution of the eigenstates, from which we determine the positions of the wave packet's maxima in both \textbf{k} space and real space. For completeness, we consider both Gaussian wave-packet and wave-packet with general distributions.

\subsection{Gaussian wave-packet}\label{litic}
Before starting with our derivation, we remind the reader that, in non-Hermitian systems, the Hamiltonian has distinct eigenfunctions for the right and left eigenvectors, $H_\textbf{k} | \psi_R^i(\textbf{k})\rangle= E^i_\textbf{k} | \psi_R^i(\textbf{k})\rangle$, $ H_\textbf{k}^\dagger | \psi_L^i(\textbf{k})\rangle = (E^i_\textbf{k})^*  | \psi_L^i(\textbf{k})\rangle$, where the eigenvectors satisfy the biorthogonality condition $\langle \psi_L^i(\textbf{k}) | \psi_R^j(\textbf{k})\rangle=\delta_{ij}$ and the generalized completeness relations $\sum_i|\psi_R^i(\textbf{k})\rangle\langle \psi_L^i(\textbf{k}) |=\sum_i|\psi_L^i(\textbf{k})\rangle\langle \psi_R^i(\textbf{k}) | =I$~\cite{Brody_2014,PhysRevX.9.041015}.

Let us now assume that the right eigenstate at the initial time $t=0$ is described by a Gaussian distribution,
\begin{equation}    |\Psi(\textbf{k},0)\rangle=|\psi_R(\textbf{k})\rangle\frac{1}{2 \pi \Delta^2} \exp \left[-\frac{1}{2}\left(\frac{\textbf{k}-\bar{\textbf{k}}}{\Delta}\right)^2\right],
\end{equation}
where $\bar{\textbf{k}}$ is the average wave vector and $\Delta$ is the Gaussian width. To avoid clutter, we avoid indicating the band label. Then, the state at time $t$ is given by
\begin{equation}
    \begin{aligned}
       |\Psi(\textbf{k},t)\rangle &=|\psi_R(\textbf{k})\rangle\frac{1}{2 \pi \Delta^2} e^{-\frac{1}{2}\left(\frac{\textbf{k}-\bar{\textbf{k}}(t)}{\Delta}\right)^2}e^{-\mathrm{i} \int_0^t d t^{\prime} E_{\textbf{k}(t^{\prime})}}e^{\mathrm{i}\gamma_\textbf{k}(t)},\\
       \bar{\textbf{k}}(t)&=\bar{\textbf{k}}(0)+\textbf{F}t,\label{eee}
\end{aligned}
\end{equation}
where $\gamma_{\textbf{k}}(t)$ is the Berry phase
\begin{equation}
    \gamma_{\textbf{k}}(t)=\int_{\textbf{k}(0)}^{\textbf{k}(t)} d \textbf{k}^{\prime} \cdot \textbf{A}_\textbf{k},
\end{equation}
with the Berry connection $\textbf{A}_\textbf{k}=\mathrm{i}\left\langle \psi_L(\textbf{k}) \mid \nabla_{\textbf{k}} \psi_R(\textbf{k})\right\rangle$~\cite{Sun1993}, that is in general complex valued.  Moreover, $E_{\textbf{k}}$ is the dispersion relation and $\mathbf{F}$ any possible external force. 

Assuming that $|\psi_R(\textbf{k}) \rangle$ varies slowly with $\textbf{k}$ on the scale of $\Delta$, the Gaussian wave-packet in real space is given by:
\begin{equation}
    \begin{aligned}
|\phi(\textbf{r},t) \rangle&= |\psi_R(\bar{\textbf{k}}(t))\rangle\frac{1}{(2 \pi)^{3} \Delta^2}\int_{\textbf{k}}  \exp [g(\textbf{k}, \textbf{r}, t)]  d\textbf{k}^2, \\ 
g(\textbf{k}, \textbf{r}, t) &= \mathrm{i} \textbf{k} \cdot \textbf{r} - \mathrm{i} \int_0^t d t^{\prime} E_{\textbf{k}(t^{\prime})} - \frac{(\textbf{k}-\bar{\textbf{k}}(t))^2}{2 \Delta^2}+\mathrm{i} \gamma_\textbf{k}(t).
\end{aligned}
\end{equation}
We notice that the employed semiclassical description assumes that the lattice constant is much smaller than the wave-packet size, and the wave-packet size is much smaller than the scale which characterizes the slow spatial variation of the applied external fields~\cite{Girvin_Yang_2019}. This hierarchy of scales defines the regime of validity of our results.

To find the maximum position of the wave-packet in real and $\textbf{k}$ space, we apply a similar strategy as in Ref. \cite{muschietti1993}. In particular, we deform the integration path into the complex $\textbf{k}$ plane and impose $\nabla_\textbf{k} g(\textbf{k}, \textbf{r}, t)=0$ to find the maximum in $\textbf{k}$ space, $\textbf{k}_{\text{M}}$. By following this procedure, we obtain the following condition,
\begin{equation}
  \textbf{r}(t)- \int_0^t d t^{\prime} \nabla_\textbf{k} E_{\textbf{k}}|_{\textbf{k}_{\text{M}}(t^\prime)}-\mathrm{i} \frac{\bar{\textbf{k}}(t)-\textbf{k}_{\text{M}}}{\Delta^2}+\nabla_\textbf{k} \gamma_\textbf{k}(t)|_{\textbf{k}_{\text{M}}}=0.
\end{equation}
Additionally, we also impose $\nabla_\textbf{r}\operatorname{Re}(g)=0$ to find the maximum in real space,  $\textbf{r}_{\text{M}}$. The solution of this condition is simply $\operatorname{Im}\textbf{k}=0$. Then, we resort to the identity \cite{Chong2010},
\begin{equation}
    \frac{d \nabla_\textbf{k}\gamma_\textbf{k}(t)}{dt}=\textbf{F}  \times \textbf{B}_\textbf{k},\label{Eq.identity}
\end{equation}
where $\textbf{B}_{\textbf{k}}=\nabla \times \textbf{A}_{\textbf{k}}$ is the Berry curvature. 

All in all, we obtain the semiclassical EOM for a non-Hermitian Gaussian wave-packet:
\begin{align}
\dot{\textbf{r}}_{\text{M}}&=\operatorname{Re}(\nabla_\textbf{k} E_{\textbf{k}}|_{\textbf{k}_{\text{M}}})- \operatorname{Re}(\textbf{F}  \times \textbf{B}_{\textbf{k}}|_{\textbf{k}_{\text{M}}}),\label{eq:eom_bulk_r}\\ \label{eq:eom_bulk_k}\dot{\textbf{k}}_{\text{M}}&=\textbf{F}+\Delta^2 [\operatorname{Im}(\nabla_\textbf{k} E_\textbf{k}|_{\textbf{k}_{\text{M}}})- \operatorname{Im}(\textbf{F}  \times \textbf{B}_\textbf{k}|_{\textbf{k}_{\text{M}}})],
\end{align}
that is one of the main results of our work.

We notice that, in Eq. \eqref{eq:eom_bulk_k}, the external force $\textbf{F}$ is accompanied by two additional contributions related to $\operatorname{Im}(E_\textbf{k})$ and $\operatorname{Im}(\textbf{B}_\textbf{k})$. These additional terms can be thought as effective forces arising because of the non-Hermitian nature of our topological system. The first of them, $\Delta^2\operatorname{Im}(\nabla_\textbf{k} E_\textbf{k}|_{\textbf{k}_{\text{M}}})$, describes local changes in dissipation or amplification, and relates to the emergent non-unitary time evolution (non-Hermiticity). This term tends to drive the system towards the $\mathbf{k}$ point with the largest $\operatorname{Im}(E_\textbf{k})$. Interestingly, the second effective force in Eq. \eqref{eq:eom_bulk_k}, $-\Delta^2 \operatorname{Im}(\textbf{F}  \times \textbf{B}_\textbf{k}|_{\textbf{k}_{\text{M}}})$, arises from geometric properties and is perpendicular to the external force. Finally, we notice that the term related to the Berry curvature exists whenever inversion or time-reversal symmetry is broken \cite{Girvin_Yang_2019}. We also stress that both the energy $E_\textbf{k}$ and the Berry curvature $\textbf{B}_\textbf{k}$ are real in the Hermitian case. Hence, the corrections in \eqref{eq:eom_bulk_k} vanish identically, leaving only the anomalous velocity term in Eq. \eqref{eq:eom_bulk_r}.

\begin{figure}[H]
\includegraphics[width=\linewidth]{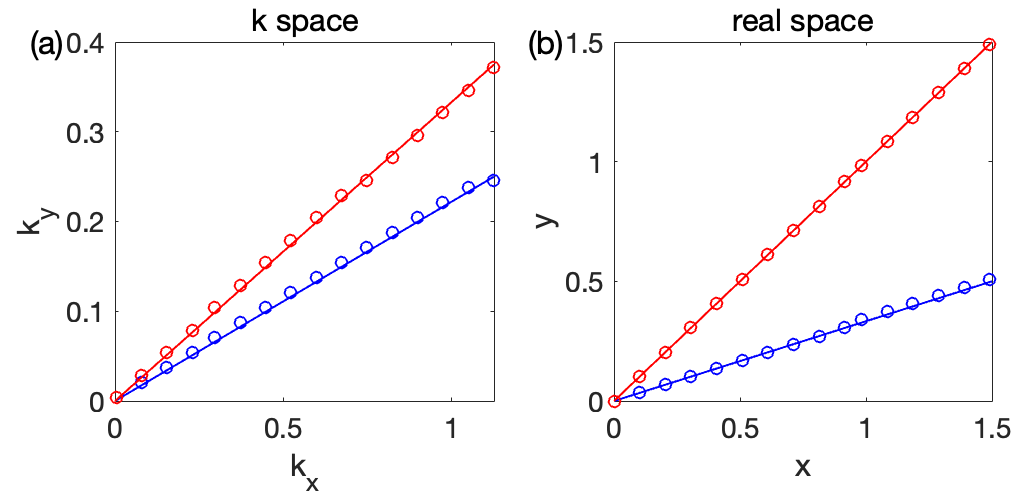}
   \caption{\textbf{2D wave-packet dynamics with and without Berry curvature:} \textbf{(a)} \textbf{k} space and \textbf{(b)} real space. We consider $B=0$ (\color{blue}blue\color{black}) and $B=1+0.5i$ (\color{red}red\color{black}). When $t=0$, we set all wave-packets centered at $(0,0)$. Then, we calculate the time evolution from $t=0$ to $t=5$ using Eq. \eqref{eee}. We use a linear dispersion $E_\textbf{k}=0.3k_x+0.1k_y+i(0.1k_x+0.2k_y)$ and take $\Delta=0.5$, $\textbf{F}=(0.2,0)$. The solid lines are calculated with Eqs.~\eqref{eq:eom_bulk_r}-\eqref{eq:eom_bulk_k} and the circles are identified with the maximum positions of the wave-packet distribution in \textbf{k} and real space, which are obtained from the numerical simulations.}
    \label{fig:linear}
\end{figure}

We test the validity of  Eqs.~\eqref{eq:eom_bulk_r}-\eqref{eq:eom_bulk_k} using numerical simulation for a 2D wave packet. For simplicity, we consider a linear (but complex) dispersion $E_{\textbf{k}}\sim k$ and a Berry connection of the form $\textbf{A}_\textbf{k}=(-B k_y,0)$. We also assume that $F_y=0$ and that $B$ is a constant, such that $\textbf{B}=\nabla \times \textbf{A}_\textbf{k}= B\hat{z}$. Thus, from Eq.~\eqref{Eq.identity}, we have  $\frac{d \nabla_\textbf{k}\gamma_\textbf{k}(t)}{dt}=(0,-BF_x)$. In the simulation, we calculate $ |\Psi(\textbf{k},t)\rangle$ in \textbf{k} space and find the position of its maximum as a function of time numerically. Then, we take the Fourier transform and locate the position of the maximum in real space as well. In Fig. \ref{fig:linear}, we show a comparison between the numerical simulations (open circles) and the predictions from our EOM, Eqs.~\eqref{eq:eom_bulk_r}-\eqref{eq:eom_bulk_k} for an arbitrary choice of $E_{\textbf{k}}$ and $B$. To confirm the validity of the EOM in non-hermitian systems, we take both quantities to be complex valued (more details are provided in the caption of Fig. \ref{fig:linear}).

Finally, we notice that Eqs.~\eqref{eq:eom_bulk_r}-\eqref{eq:eom_bulk_k} hold for both 2D and 3D systems. On the other hand, the EOM for a 1D wave-packet are given by:
    \begin{align}
\dot{x}_{\text{M}}&=\operatorname{Re}\left(\frac{d E_{k}}{dk}|_{k_{\text{M}}}\right),\label{eq:eom_edge0}\\ \dot{k}_{\text{M}}&=F+\Delta^2 \operatorname{Im}\left(\frac{d E_{k}}{dk}|_{k_{\text{M}}}\right), \label{eq:eom_edge}
\end{align}
and they correspond to the 1D limit of Eqs.~\eqref{eq:eom_bulk_r}-\eqref{eq:eom_bulk_k}, \textit{i.e.}, setting $\frac{d^2 \gamma_k(t)}{dkdt}=0$.

\subsection{Wave-packet dynamics beyond the Gaussian limit}
Our analytic derivation of the wave-packet EOM in Section~\ref{litic} implicitly assumes that the wave-packet has a Gaussian shape and that this shape is preserved during the whole time evolution. However, if $\operatorname{Im}(\nabla_{\textbf{k}} E_\textbf{k})$ and $\nabla_{\textbf{k}} \gamma_\textbf{k}$ are not constant in $\textbf{k}$, or equivalently $\operatorname{Im}(E_\textbf{k})$ and $\gamma_\textbf{k}$ are nonlinear functions of $\textbf{k}$, the Gaussian shape in $\textbf{k}$ space will not be preserved under the time evolution since different $\textbf{k}$ component will move at different speeds. In other words, the previously derived equations are valid only in the limit where the dispersion and the Berry curvature are approximately linear in $\textbf{k}$.

In a more general situation, the wave-packet shape evolves with time and deviates from the Gaussian form. In order to capture this more general case, we extend the previous analysis by assuming an undetermined distribution in $\textbf{k}$ space $\exp\left[ -f(\textbf{k},t,\textbf{F}) \right]$, which is in general a function of time and the external force $\textbf{F}$ (to be compared with the simpler Gaussian form assumed in Eq.~\eqref{eee}).

Following the same logic as in Section~\ref{litic}, the time evolved wave-packet in wave vector space takes the following form:
\begin{equation}
    |\Psi(\textbf{k},t)\rangle =|\psi_R(\textbf{k})\rangle\exp\left[ -f(\textbf{k},t,\textbf{F}) \right] e^{-i \int_0^t d t^{\prime} E_{\textbf{k}(t^{\prime})}} e^{i \gamma_\textbf{k}(t)}.
\end{equation}
At the same time, the real space wave-packet follows
\begin{align}
&|\phi(\textbf{r},t)\rangle=|\psi_R(\bar{\textbf{k}}(t))\rangle\frac{1}{2 \pi}\int_\textbf{k}  \exp [g(\textbf{k}, \textbf{r}, t)]  dk,\nonumber\\ 
     &g(\textbf{k}, \textbf{r}, t) \equiv -f(\textbf{k},t,\textbf{F}) + i \textbf{k} \cdot \textbf{r}-i \int_0^t d t^{\prime} E_{\textbf{k}(t^{\prime})}+i \gamma_\textbf{k}(t).
\end{align}
Continuing with the derivation, we assume as before that $\nabla_\textbf{k} g|_{\textbf{k}_{\text{M}}}=0$ and further apply a time derivative. Then, we obtain the final equations describing the wave-packet dynamics,
\begin{align}
    &\dot{\textbf{r}}_{\text{M}}=\operatorname{Re}(\nabla_\textbf{k} E_{\textbf{k}}|_{\textbf{k}_{\text{M}}})- \operatorname{Re}(\textbf{F}  \times \textbf{B}_{\textbf{k}}|_{\textbf{k}_{\text{M}}})\label{eq:eom_general1},\\
        &\frac{d \nabla_\textbf{k} f(\textbf{k},t,\textbf{F})}{d t}|_{\textbf{k}_{\text{M}}}= \operatorname{Im}\left(\nabla_\textbf{k}E_{\textbf{k}}|_{\textbf{k}_{\text{M}}}\right)-\operatorname{Im}(\textbf{F}  \times \textbf{B}_\textbf{k}|_{\textbf{k}_{\text{M}}}).\label{eq:eom_general2}
\end{align}
We notice that the non-Gaussian shape of the wave-packet does not affect the equation for $\textbf{r}_{\text{M}}$, Eq.~\eqref{eq:eom_general1}. However, non-Gaussianities modify the dynamics of $\textbf{k}_{\text{M}}$, introducing a dependence on the distribution function $f$ rather than a direct relationship with $\dot{\textbf{k}}_{\text{M}}$ as before. Moreover, the rotational invariance in the $\textbf{k}$ space can now be explicitly broken, depending on the choice of $f$. Eq.~\ref{eq:eom_general2} reduces to Eq.~\ref{eq:eom_bulk_k} when we set the distribution to be Gaussian, \textit{i.e.},
\begin{equation}
    f_{\text{Gaussian}}(\textbf{k},t,\textbf{F})=-\frac{1}{2}\left(\frac{\textbf{k}-\bar{\textbf{k}}(t)}{\Delta}\right)^2.
\end{equation}

\section{Topological wave-packet dynamics in Non-Hermitian Haldane model}
In this section, we calculate the complex Berry curvature, Chern number, and chiral edge states for a non-Hermitian Haldane model with non-reciprocal NNN hopping terms. We then investigate the bulk-edge correspondence and present numerical simulations of wave-packet dynamics.
 
\subsection{Non-Hermitian Haldane model}\label{mod}
\begin{figure}[H]
\includegraphics[width=\linewidth]{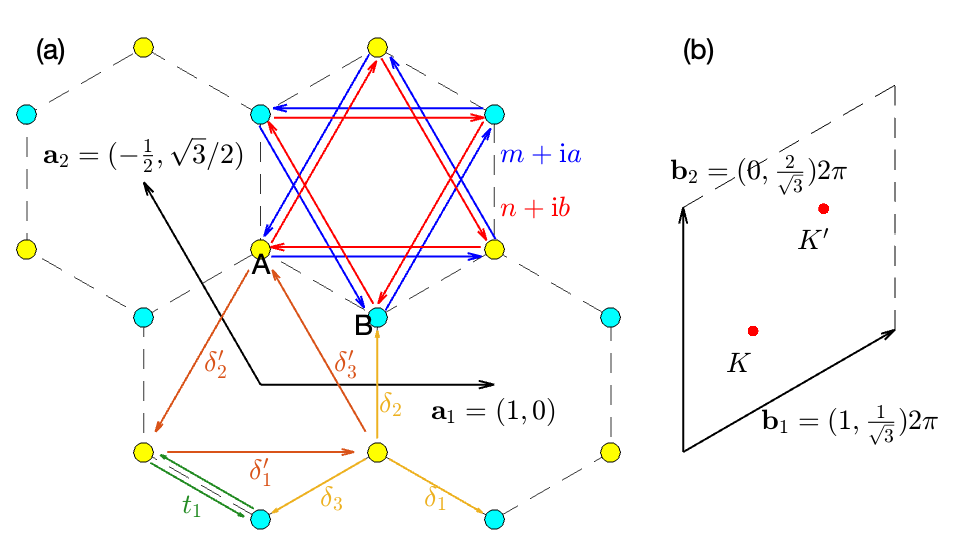}
    \caption{\textbf{The model considered in this work: }\textbf{(a)} Honeycomb lattice with A (yellow) and B (cyan) sublattices. The NNN hopping parameters are indicated with blue and red arrows. The NN hopping parameter $t_1$ is represented with green arrows. \textbf{(b)} The first Brillouin zone, where the wave vector is expressed as $\textbf{k}=k_1 \textbf{b}_1 +k_2 \textbf{b}_2$. The high-symmetry points $K=(\frac{1}{3},\frac{1}{3})$ and $K^\prime=(\frac{2}{3},\frac{2}{3})$ are also indicated with red dots.}
    \label{fig:1}
\end{figure}

We consider an inversion symmetric non-Hermitian honeycomb lattice with nearest-neighbor (NN) hopping $t_1$ and complex NNN hoppings. The NNN `anticlockwise' hoppping parameter is given by $m+\mathrm{i}a$, while the `clockwise' one is given by $n+\mathrm{i}b$. The Hermitian limit in this model could be recovered by imposing $m=n$ and $a=-b$. We notice that PT symmetry is always broken, hence the eigenvalues of the Hamiltonian are expected to be complex. The main features of the model are illustrated in Fig.~\ref{fig:1} 
The tight-binding Hamiltonian in $\textbf{k}$ space is given by: 
\begin{equation}
    H_\textbf{k}= \left(\begin{array}{cc} g_\textbf{k}+p_\textbf{k}& f_\textbf{k}\\ f_\textbf{k}^* & g_\textbf{k}-p_\textbf{k} \end{array}\right).
\end{equation}
Here, we define

\begin{align}
    &f_\textbf{k}= t_1\sum_j e^{\mathrm{i}\mathbf{k}\cdot \bm{\delta_j}},\nonumber\\
    &g_\textbf{k}=[m+n+\mathrm{i}(a+b)] \sum_j\cos(\mathbf{k}\cdot  \bm{\delta}_j^\prime),\nonumber\\
    &p_\textbf{k}=[(a-b)+\mathrm{i}(n-m)]\sum_j\sin(\mathbf{k}\cdot  \bm{\delta}_j^\prime)],
\end{align}
where $\bm{\delta}_1=\frac{1}{3} \mathbf{a}_1-\frac{1}{3} \mathbf{a}_2$, $\bm{\delta}_2= \bm{\delta}_1+ \mathbf{a}_2$, and $\bm{\delta}_3= \bm{\delta}_1- \mathbf{a}_1$ are the vectors describing the NN hopping terms. Moreover $\bm{\delta}_1^\prime= \mathbf{a}_1$, $\bm{\delta}_2^\prime= -\mathbf{a}_1-\mathbf{a}_2$, and $\bm{\delta}_3^\prime= \mathbf{a}_2$
are the vectors representing the NNN hopping terms. All these vectors are shown explicitly in Fig.~\ref{fig:1}. For simplicity, $t_1$ is set to be $1$ in this work. For later convenience, we define the ``complex chirality" of the real and imaginary parts of the NNN hoppings as
\begin{align}
    &S_{\R}=\sign \left(n-m\right),\quad S_\I=\sign \left(b-a\right).
\end{align}

\begin{figure}[h]
\includegraphics[width=\linewidth]{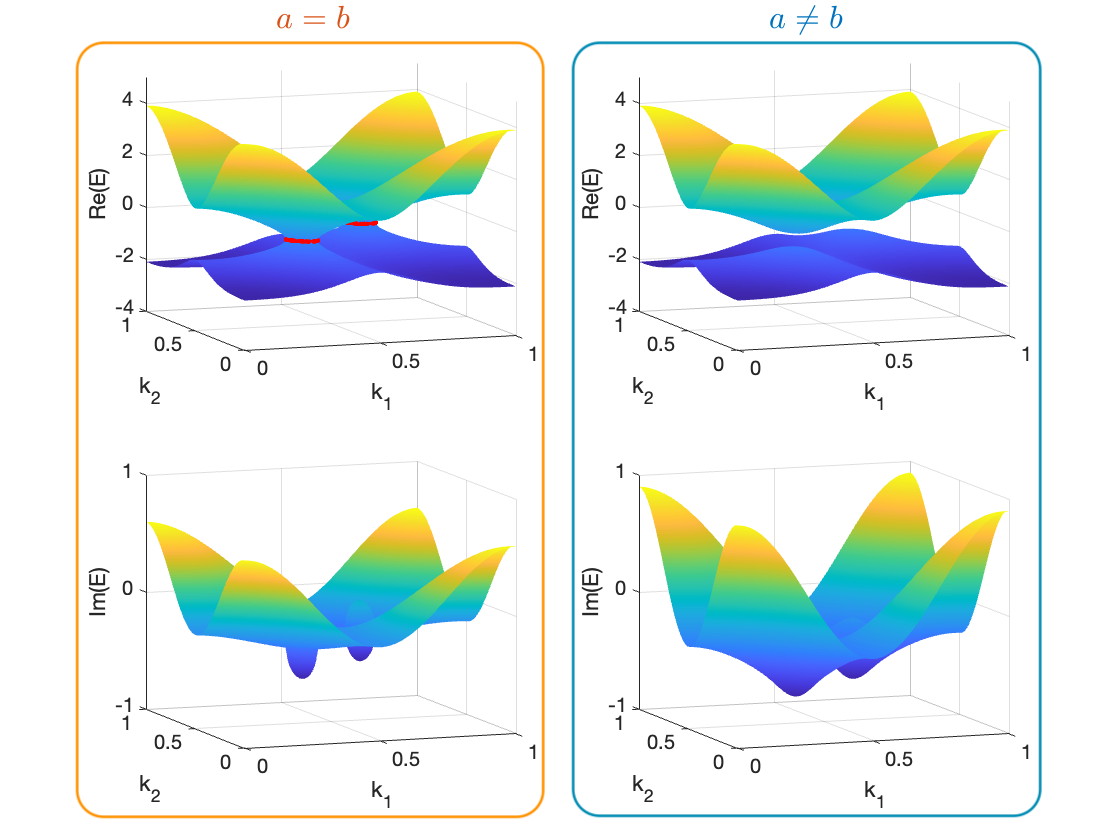}
\caption{\textbf{Band dispersion: } \textbf{Left column:} real and imaginary energy spectrum for $a=b=0.1$. The two red rings mark the exceptional rings, where both the real and imaginary parts of the energy become degenerate. \textbf{Right column:} real and imaginary energy spectrum for $a\neq b$, $a=0.1$ and $b=0.2$, with real gap opened. In all these figures, $m=0.1$ and $n=0.2$.}
    \label{fig:2}
\end{figure}

The system exhibits inversion symmetry, where the inversion centers correspond to the center of each `honeycomb' or the points in between the NN sites. The tight-binding Hamiltonian satisfies $H_\textbf{k}=\sigma_x H_{-\textbf{k}}\sigma_x$~\cite{Bernevig+2013}, with $\sigma$ being the Pauli matrix.

\begin{figure*}[ht!]
\includegraphics[width=\linewidth]{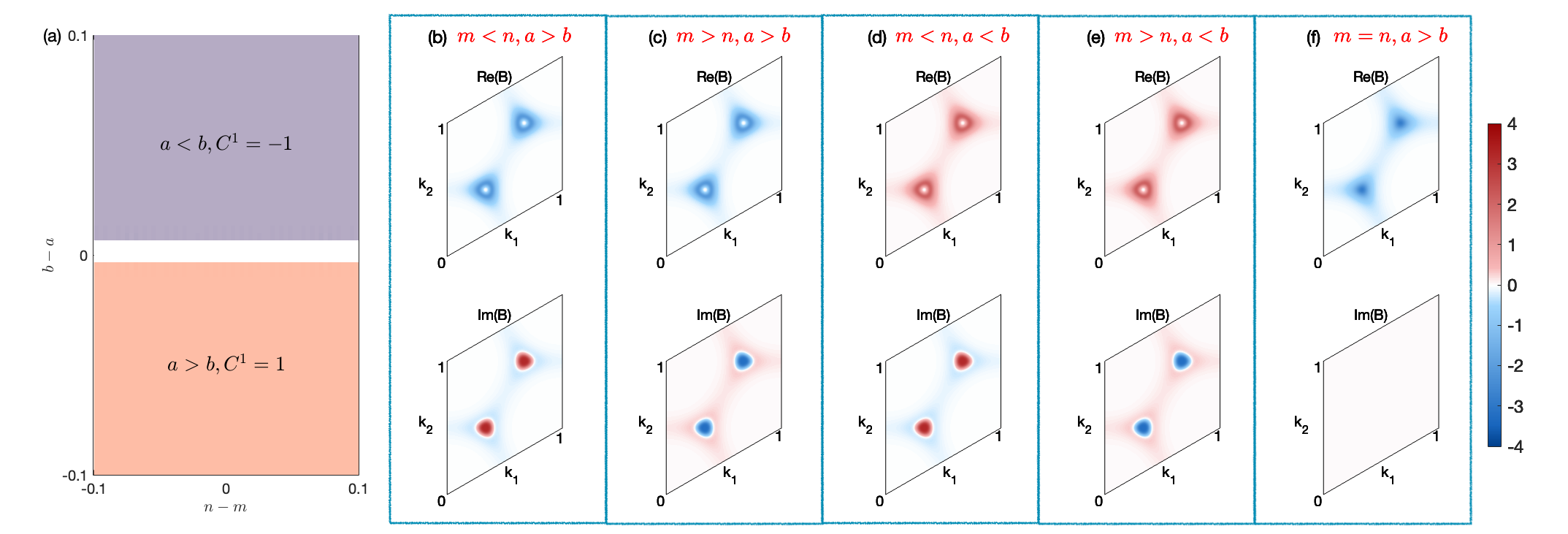}
    \caption{\textbf{Chern number and Berry curvature: }\textbf{(a)} Dependence of the Chern numbers on the NNN hopping parameters. \textbf{(b-f)} Real and imaginary part of lower-band Berry curvature for different hopping parameters: \textbf{(b)} $m$ = 0.1, $n$ = 0.2, $a$ = 0.2, $b$= 0.1; \textbf{(c)} $m$ = 0.2, $n$ = 0.1, $a$ = 0.2, $b$= 0.1; \textbf{(d)} $m$ = 0.1, $n$ = 0.2, $a$ = 0.1, $b$= 0.2; \textbf{(e)} $m$ = 0.2, $n$ = 0.1, $a$ = 0.1, $b$= 0.2; \textbf{(f)} $m$ = 0.1, $n$ = 0.1, $a$ = 0.25, $b$= 0.1;}
    \label{fig:3}
\end{figure*}

\begin{figure*}[t]
\includegraphics[width=\linewidth]{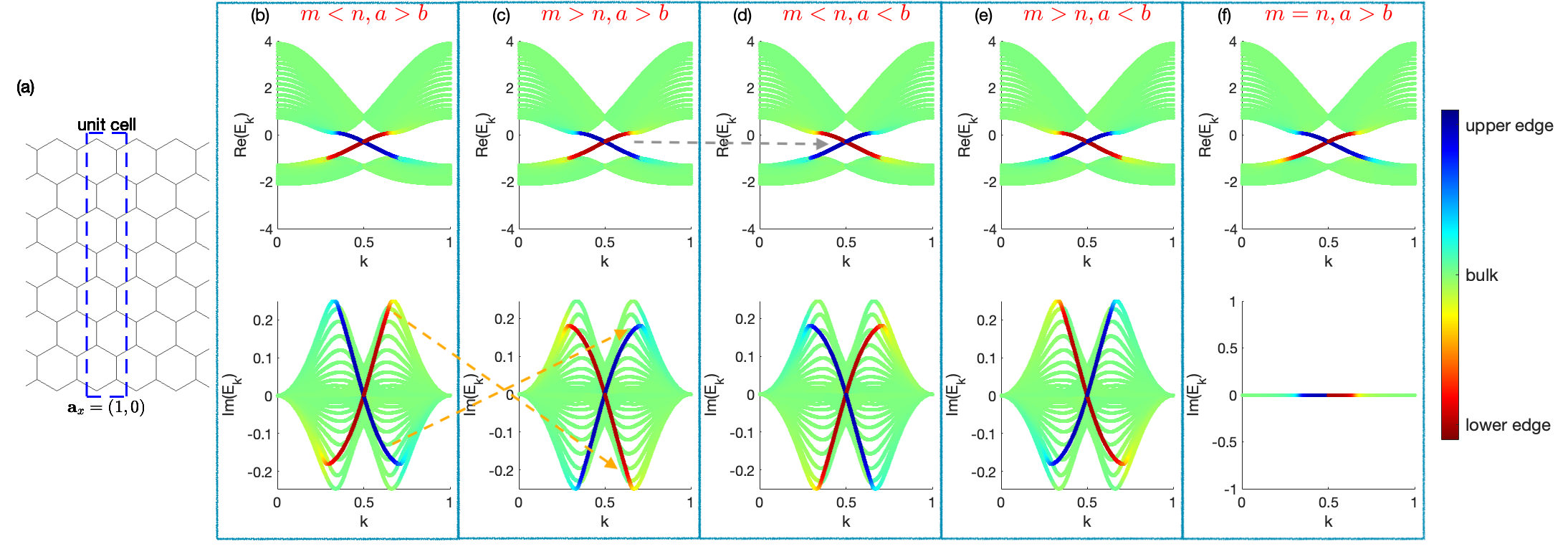}    \caption{\textbf{Chiral edge states on zigzag edges: }\textbf{(a)} An illustration of the honeycomb lattice strip with zigzag edge. \textbf{(b-f)} Real and imaginary part of the energy spectrum for the strip with different hopping parameters: \textbf{(b)} $m$ = 0.1, $n$ = 0.2, $a$ = 0.1, $b$= -0.1; \textbf{(c)} $m$ = 0.2, $n$ = 0.1, $a$ = 0.1, $b$= -0.1; \textbf{(d)} $m$ = 0.1, $n$ = 0.2, $a$ = -0.1, $b$= 0.1; \textbf{(e)} $m$ = 0.2, $n$ = 0.1, $a$ = -0.1, $b$= 0.1; \textbf{(f)} $m$ = 0.15, $n$ = 0.15, $a$ = 0.1, $b$= -0.1; We set 40 atoms in one unit cell for the calculation. The blue/red energy bands represent eigenstates localized on the upper/lower edge of the strip, while the other (green) states are bulk states. In all figures, we have set $a=-b$. The gray dashed arrow illustrates the inversion of the edge state chirality by changing the sign of $S_\I$. The orange dashed lines indicate the inversion of the $\operatorname{Im}( E_\textbf{k})$ of the chiral edge states with respect to the horizontal axes upon changing the sign of $S_\R$.}
    \label{fig:4}
\end{figure*}

The complex dispersion relation of the two electronic bands is given by
\begin{equation}
    E_{\textbf{k},\pm}=g_\textbf{k}\pm \sqrt{p_\textbf{k}^2+f_\textbf{k}f_\textbf{k}^*}
\end{equation}
and it is shown in Fig.~\ref{fig:2} in the two cases $a=b$ (or equivalently $S_\I=0$) and $a \neq b$.
When $S_\I=0$, both the real and imaginary parts of $ E_{\textbf{k},\pm}$ are degenerated at $f_\textbf{k}f_\textbf{k}^*=[\text{Im}(p_\textbf{k})]^2$. This defines the so-called `exceptional rings' around $K$ and $K^\prime$, as shown in the left column of Fig.~\ref{fig:2} by red solid lines. These exceptional rings shrink to degenerated points, $K$ and $K^\prime$, when $S_\R=S_\I=0$. On the other hand, when $a \neq b$ (right column of the same figure), a topologically protected gap between the two bands appears.

\subsection{Berry curvature and Chern number}

The topological invariant Chern number for the $i$th band is defined by the integral of the Berry curvature $\textbf{B}^i_{\textbf{k}}$ over the first Brillouin zone~\cite{asboth2016short,PhysRevB.84.205128}:
\begin{equation}
    C^i=-\frac{1}{2\pi}\int_{BZ} (\textbf{B}^i_{\textbf{k}})^z d^2k, \label{defC}
\end{equation}
where
\begin{equation}
\begin{aligned}
\textbf{B}^i_{\textbf{k}}&=\nabla \times \textbf{A}^i_{\textbf{k}}\\ &=\mathrm{i} \sum_{j \neq i} \frac{\langle \psi_L^i(\textbf{k})|\nabla H_{\textbf{k}}| \psi_R^j(\textbf{k}) \rangle \times\langle \psi_L^j(\textbf{k})|\nabla H_{\textbf{k}}| \psi_R^i(\textbf{k})\rangle}{\left(E^i_{\textbf{k}}-E^j_{\textbf{k}}\right)^2}.
\end{aligned}\label{eq:Berry}
\end{equation}

Following the definition in Eq.~\eqref{defC}, we compute the Chern numbers $C^i$ in our non-Hermitian model defined in the previous section. In general, the Chern numbers are non-trivial functions of $S_\I$. When $S_\I=0$, the energy gap closes, and the Chern number becomes ill-defined. When $S_\I\neq 0$, the time-reversal symmetry, $H_\mathbf{k}=H_\mathbf{-k}^*$ and $\textbf{B}^i_{\textbf{k}}=-\textbf{B}^i_{-\textbf{k}}$~\cite{Girvin_Yang_2019}, is explicitly broken, allowing for non-zero Chern numbers. In panel (a) of Fig.~\ref{fig:3}, we show the value of the Chern number of the lower (first band), $C^{1}$, as a function of $n-m$ and $b-a$, describing the NNN hopping parameters. We observe that $C^1=-1$ when $a<b$ and $C^1=1$ when $a>b$, while it is insensitive to the sign of $n-m$. This indicates that:
\begin{equation}
    C^{1}= - S_\I.
\end{equation}
In order to characterize in more detail the topological properties of our model, in Fig.~\ref{fig:3} we show the real and imaginary parts of the Berry curvature, $\operatorname{Re}(\mathbf{B})$ and $\operatorname{Im}(\mathbf{B})$, in the first Brillouin zone for the four different scenarios $ \left(S_\I \gtrless 0\right) \wedge \left(S_\R \gtrless 0\right)$. As a general feature, the Berry curvature always peaked around the high-symmetry points $K, K'$ where its value grows by closing the gap in the electronic bands. 

More importantly, as shown in panels (b-f) of Fig.~\ref{fig:3}, the sign of $\operatorname{Re}(\mathbf{B})$ is proportional to $S_\I$. Moreover, although the integral of the imaginary Berry curvature over the first Brillouin zone identically vanishes, the sign of $\operatorname{Im}(\textbf{B})$ (see panels (b-f) of Fig.~\ref{fig:3}) is proportional to $S_\R$. Finally, when $S_\R=0$, the imaginary Berry curvature is identically zero, even when the system is non-Hermitian ($a\neq -b$).

It is interesting to notice that $S_\I$ controls the real part of the Berry curvature, while $S_\R$ determines the imaginary part of the Berry curvature. This structure deserves further study.

In Appendix~\ref{app.A}, we discuss further topological properties of this model related to the quantum geometric tensor.

\subsection{Chiral edge states on zigzag edges and bulk-edge correspondence}\label{sec:bulk-edge}
After analyzing the topological properties of our system in terms of the Berry curvature and the Chern number, we are ready to connect them through the bulk-edge correspondence to the existence of chiral edge states.
In this regard, we consider a strip in honeycomb lattice with
zigzag edges in the y-direction, as shown in panel (a) of Fig.~\ref{fig:4}. 

In panels (b-f) of Fig.~\ref{fig:4}, we show the band spectrum $E_k$ with open boundary conditions corresponding to the strip shown in panel (a). The color of the various curves indicates the localization properties of the relative modes with green corresponding to bulk modes, while blue/red colors correspond to edge states localized on the upper and lower edge respectively. The degree of localization is determined by summing the modulus square of the eigenvector components with a weight that varies across the sublattices: $\sum_{j=1}^{N_u} |v_j|^2(-1+\frac{2j-2}{N_u-1})$, where $j\in [1,N_u]$ indexes the sublattices from top to bottom, and $v_j$ is $j$th component of eigenvector. For $j = 1$ and $j = N_u$, the weights are $-1$ and $1$, corresponding to the upper and lower edges, respectively.

As expected, the sum of the Chern numbers up to the $i$th band determines the number and chirality of the edge states in the $i$th band gap. For our two-band system, the edge states are governed by the Chern number of the lower band, $C^1$, which satisfies $C^1=-C^2$. More precisely, $C^{1}=1$ and $C^{1}=-1$ correspond respectively to a anticlowise/clockwise moving edge state \cite{PhysRevB.90.024412,PhysRevLett.71.3697}. From Fig.~\ref{fig:4}, we observe that the chirality of the edge states is also determined by the value of $S_\I$. When $S_\I<0$, the `anticlockwise' imaginary NNN hopping is larger than the `clockwise' one, resulting in an anticlockwise propagating topological edge state. Conversely, when $S_\I>0$, the situation is reversed, leading to a clockwise propagating edge state. This inversion is illustrated as an example in panels (c) and (d) of Fig.~\ref{fig:4} with a gray dashed arrow.

Finally, $S_\R$ controls the behavior of $\operatorname{Im}(E_k)$, which represents the gain or loss of the system as a function of $\textbf{k}$. In Fig.~\ref{fig:4}, we set $a=-b$ so that the Hamiltonian is Hermitian and $\operatorname{Im}(E_k)=0$ when $m=n$. Then, the shape of $\operatorname{Im}(E_k)$ flips relative to the horizontal axes, $\operatorname{Im}(E_k)$=0,  when $S_\R$ changes sign, as indicated in panels (b-c) of Fig.~\ref{fig:4} with orange dashed arrows. This mirror transformation is also accompanied by a change in the sign of $\operatorname{Im}(\textbf{B})$. When $a\neq -b$, $\operatorname{Im}(E_k)$ does not simply change sign with $S_\R$, but $S_\R$ still influences the behavior of $\operatorname{Im}(E_k)$. A similar analysis of the topologically protected edge states in the case of armchair edges is presented in Appendix~\ref{app.B}.

To summarize, the real part of Berry curvature acts as an anomalous contribution to the velocity of the bulk wave-packet and determines the velocity of the edge wave-packet. Furthermore, the imaginary part of the Berry curvature induces an anomalous force to the bulk wave-packet. The anomalous force for the bulk wave-packet and the effective force $\Delta^2 \operatorname{Im}(\frac{d E_{k}}{dk}|_{k_{\text{M}}})$ for the edge wave-packet change sign with $S_\R$. This suggests the existence of a novel correspondence that deserves further attention.

\subsection{Edge wave-packet dynamics}

\begin{figure}[h]
\includegraphics[width=\linewidth]{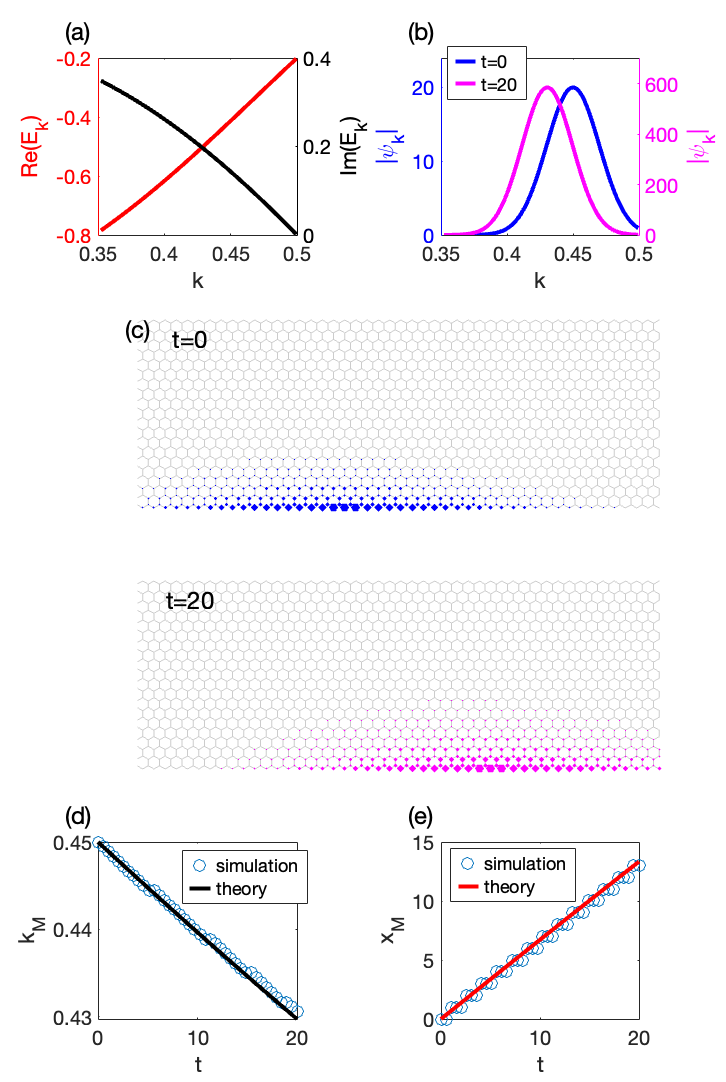}
    \caption{\textbf{Edge wave-packet dynamics:} we choose the edge state localized on the lower edge and for simplicity set the external force to zero. \textbf{(a)} Energy spectrum $E_{\textbf{k}}$ used in the simulation. We fix $m = 0.2$, $n = 0.1$, $a = 0.1$, $b= -0.1$. \textbf{(b)} Propagation of the wave-packet in $\textbf{k}$ space. The initial state has maximum at $k$ = 0.45 and a width of $\Delta$ = 0.02.  \textbf{(c)} Propagation of wave-packet in real space. The initial state has maximum at $x$ = 0. \textbf{(d, e)} Comparison of the simulation results and the theoretical predictions, obtained from Eqs.~\eqref{eq:eom_edge0}-\eqref{eq:eom_edge}, for the maxima of the edge wave-packet in $k$ and real space, $k_{\text{M}}$ and $x_{\text{M}}$.}
    \label{fig:5}
\end{figure}

We focus on the numerical simulations for the dynamics of the edge wave-packet, Eqs.~\eqref{eq:eom_edge0}-\eqref{eq:eom_edge}. In the simulation, we choose the range of the strip's energy in the middle of the gap in the bulk spectrum, corresponding to the energy range where the edge states exist. The edge energy spectrum with positive $\operatorname{Im}(E_k)$ used in the calculation is shown in panel (a) of Fig.~\ref{fig:5}.

We then calculate the dynamics of the wave-packet in the 1D $k$ space, $\frac{1}{\sqrt{2 \pi} \Delta} e^{-\frac{1}{2}\left(\frac{k-\bar{k}}{\Delta}\right)^2}e^{-\mathrm{i} E_k t}$, as a function of time. Two snapshots of the initial wave-packet and the wave-packet at later time ($t=20$) are shown in panel (b) of Fig.~\ref{fig:5}. The wave-packet in $k$ space moves to smaller $k$ because $\operatorname{Im}(\frac{d E_{k}}{dk}|_{k_{\text{M}}})$ is negative. The total amplitude increases due to the positive value of $\operatorname{Im}(E_k)$. 

Then, we take a Fourier transform along the $x$ axis and use the eigenfunction spread along the $y$ direction to calculate the wave-packet in real space, as shown in panel (c) of Fig.~\ref{fig:5}. The wave-packet in real space moves to the right because $\operatorname{Re}(\frac{d E_{k}}{dk}|_{k_{\text{M}}})$ is positive. 

Finally, in panels (d) and (e) of Fig.~\ref{fig:5}, we show with open symbols the time dependence of the wave-packet maxima in $k$ and real space, respectively $k_{\text{M}}$ and $x_{\text{M}}$. In the same figures, with solid black and red lines, we show the predictions from the theory obtained using Eqs.~\eqref{eq:eom_edge0}-\eqref{eq:eom_edge}. The agreement between the simulations and the theoretical estimate is excellent, confirming the validity of Eqs.~\eqref{eq:eom_edge0}-\eqref{eq:eom_edge}. 
 
 \subsection{Bulk wave-packet dynamics}

\begin{figure}[h]
\includegraphics[width=\linewidth]{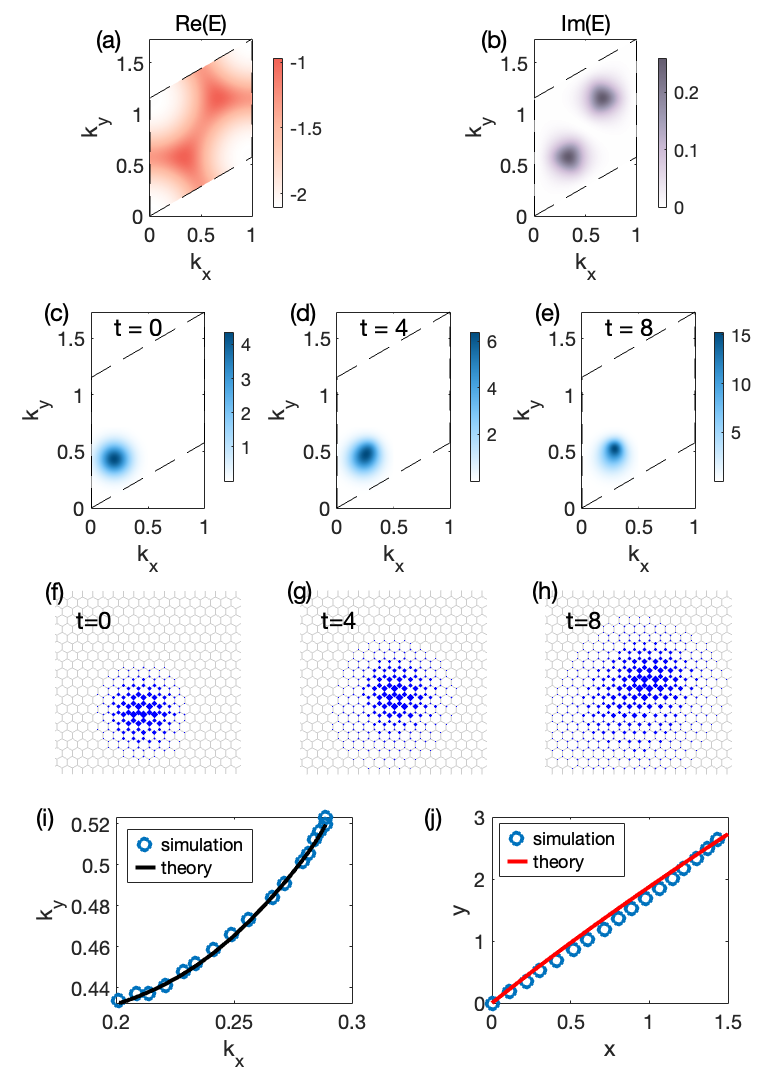}
    \caption{\textbf{Bulk wave-packet dynamics:} Bulk wave-packet simulation using the same parameters as Fig.~\ref{fig:5} with force $\Vec{F}=(0.05,0)$; \textbf{(a-b)} Energy spectrum used in the simulation. We fix $m = 0.2$, $n = 0.1$, $a = 0.1$, $b= -0.1$. The first Brillouin zone is indicated by dashed lines. \textbf{(c-e)} Propagation of the bulk wave-packet in $\textbf{k}$ space. The initial state has maximum at $k_x=0.20$, $k_y=0.43$ and $\Delta=0.1.$ \textbf{(f-h)} Propagation of the bulk wave-packet in real space. The initial state has a maximum at $x=y=0$.\textbf{(i-j)} Comparison between the simulation results and the theoretical predictions using Eqs.~\eqref{eq:eom_general1}-\eqref{eq:eom_general2} for the position of the bulk wave-packet in $k$ space and real space. The position is identified with the location of the maximum of the wave-packet distribution.}.
    \label{fig:6}
\end{figure}

For our system, the bulk spectrum is not linear in $\textbf{k}$ (panels (a) and (b) in Fig.~\ref{fig:6}), and does not display isotropy in the $k_x,k_y$ plane. Therefore, the more general Eqs.~\eqref{eq:eom_general1}\eqref{eq:eom_general2} have to be considered. 

In our simulation, we use the spectrum shown in panels (a)-(b) of Fig.~\ref{fig:6}, where the band with positive $\operatorname{Im}E_{\textbf{k}}$ has been selected. We then calculate the time evolution of the wave-packet in 2D $\textbf{k}$ space. The corresponding dynamics in real space can be directly obtained via a Fourier transform.

In panels (c)-(e), we show three snapshots of the time evolution of the bulk wave-packet in $\textbf{k}$ space from the initial state at $t=0$ to $t=8$. As evident from there, despite the initial distribution is isotropic and Gaussian, at later times the shape becomes anisotropic and non Gaussianities emerge. A similar behavior is found for the time evolution of the real space bulk wave-packet, that is shown in panels (f)-(h) of Fig.~\ref{fig:6} in the same interval of time.

Finally, in panels (i) and (j) of Fig.~\ref{fig:6}, we compare the simulation results with the theoretical predictions by tracking the position of the wave-packet in $\textbf{k}$ space and in real space upon evolving time. We observe a good agreement between the two. We notice that the Berry phase is not included in the simulation and theoretical calculation due to numerical challenges caused by its gauge dependence.

\section{Realization with a non-Hermitian Chiral cavity}
\label{sec:realization}
In this section, we propose a realization of the non-Hermitian Haldane model discussed above using a non-Hermitian optical chiral cavity.

We consider an Hermitian Honeycomb lattice with real hopping parameters inside a non-Hermitian chiral cavity. The Hamiltonian that captures the interaction between a charged particle and the photonic mode in the cavity is given by
\begin{equation}
\begin{aligned}
        \hat{H}&=\frac{1}{2 m}(\hat{\mathbf{p}}-q \hat{\mathbf{A}})^2+V(\mathbf{r})+\hbar \omega_c \hat{a}^{\dagger} \hat{a},\\
        \hat{\mathbf{A}}&=A_0\left(\boldsymbol{f} \hat{a}+\gamma\boldsymbol{f}^* \hat{a}^{\dagger}\right),\quad A_0=\sqrt{\frac{\hbar}{2 |\omega_c| V \epsilon_0}},\label{eq:H}
    \end{aligned}
\end{equation}
where
\begin{equation}
    \boldsymbol{f}=\frac{1}{\sqrt{2}}(1,-\mathrm{i}) \cos \theta+\frac{1}{\sqrt{2}}(1,\mathrm{i}) \sin \theta
\end{equation}
represents the superposition of right- and left-handed polarized light. Here, $\theta=0 (\pi/2)$ corresponds to right-handed (left-handed) circular polarization. A prefactor $\gamma \neq 1$ breaks hermiticity by making $\hat{\mathbf{A}}^\dagger \neq \hat{\mathbf{A}}$. In other words, for $\gamma \neq 1$, we are considering `non-Hermitian photons'. Thus $\omega_c=\frac{|\omega_c|}{\gamma}$ ($|\gamma|=1$) is complex when $\gamma$ is complex. Positive/negative $\operatorname{Im}(\gamma)$ corresponds to dissipation/amplification of the cavity, respectively. Finally, $m$ is the mass of the particle, $V(\mathbf{r})$ is an external potential.

We then use a similarity transformation $\hat{U}=\exp \left[-\mathrm{i} \frac{\xi}{\hbar} \hat{\mathbf{p}} \cdot \hat{\boldsymbol{\pi}}\right]$ to  asymptotically decouple the photon and matter degrees of freedom, where $\xi=\frac{g}{1+g^2} \sqrt{\frac{\hbar}{m \omega_c}}$. The parameter $g=\sqrt{\frac{\left(q A_0\right)^2}{m \hbar \omega_c}}$ quantifies the strength of the light-matter coupling. Conveniently, the introduced parameter $\xi$ approaches zero in both the weak/strong-coupling limits \cite{PhysRevB.107.195104,PhysRevLett.126.153603,jiang2023}. Moreover, $\hat{\bm{\pi}}=\mathrm{i}\left(\gamma\boldsymbol{f}^* \hat{a}^{\dagger}-\boldsymbol{f} \hat{a}\right)$ represents the photonic momentum operator and satisfies the commutation relation $\left[\hat{\bm{\pi}}_x, \hat{\bm{\pi}}_y\right]=i \gamma\cos(2\theta)$.

After applying the above transformation and expanding the resulting expressions to the first-order in $\xi$, we obtain the effective Hamiltonian,
\begin{align}
\hat{H}_U&=\hat{U}^{-1} \hat{H} \hat{U} \nonumber \\ &\approx \frac{(\hat{\mathbf{p}}-q \mathbf{A(\mathbf{r})})^2}{2 m_{\mathrm{eff}}}+V\left(\mathbf{r}\right)+\hbar \omega_{\mathrm{eff}} \hat{a}^{\dagger} \hat{a}+\nabla V(\mathbf{r}) \xi \hat{\pi},
\end{align}
where
\begin{align}
    &\mathbf{A(\mathbf{r})}  = \gamma\cos(2\theta) ~\frac{m_{\mathrm{eff}} \xi^2}{2 q \hbar} (\mathbf{e}_{\mathbf{z}} \times \nabla V(\mathbf{r})),\\
    &m_{\mathrm{eff}}=m\left(1+g^2\gamma\right),\\
&\omega_{\mathrm{eff}}=\omega_c\left(1+g^2\right).
\end{align}
The detailed derivation of these results is provided in Appendix~\ref{app.C}.

An inversion symmetric potential for the honeycomb lattice is then assumed as $V(\mathbf{r})=\sum_{i=1}^6V_0 e^{ \mathrm{i} \boldsymbol{n}_i \cdot \boldsymbol{r}}$, where $\boldsymbol{n}_i=\pm \mathbf{b}_1, \pm \mathbf{b}_2, \pm \left(\mathbf{b}_1-\mathbf{b}_2\right)$. The minima of the potential where the atoms sit form the honeycomb lattice. A representation of the honeycomb potential is provided in panel (a) of Fig.~\ref{fig:6} with the black spots representing the atomic positions. We then use Peierls substitution to calculate how the gauge field $\mathbf{A}(\mathbf{r})$ effectively modifies the hopping constants in the honeycomb lattice. This minimal coupling can be then implemented following
\begin{equation}
        t_{ij} \rightarrow t_{ij} e^{\mathrm{i}\phi_{ij}}, \quad \phi_{i j}=-\frac{q}{\hbar} \int_{\boldsymbol{r}_i}^{\boldsymbol{r}_j} \mathbf{A(\mathbf{r})} \cdot d \textbf{r}, \label{pp}
\end{equation}
where $t_{ij}$ are the hopping constants from site $i$ to site $j$ in the lattice.

\begin{figure}[h]
\includegraphics[width=\linewidth]{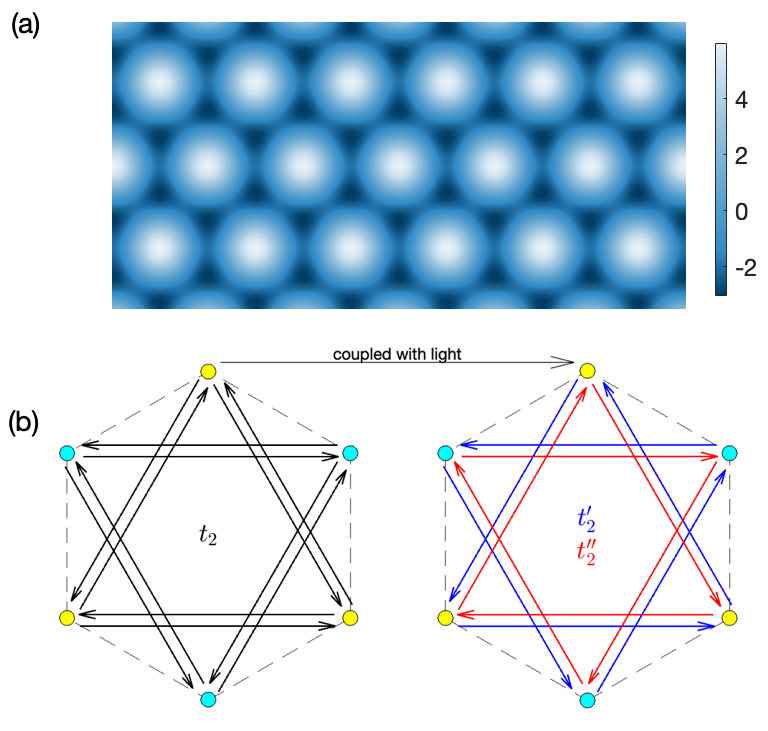}
    \caption{\textbf{(a)} Inversion symmetric potential $V(\textbf{r})$ with $V_0=1$ forming a honeycomb lattice. \textbf{(b)} An illustration of how light-matter interaction modifies the NNN hopping parameters and make the system non-Hermitian.}
    \label{fig:7}
\end{figure}

For NN hopping, a direct evaluation of Eq.~\eqref{pp} implies $\phi_{ij}=0$. On the other hand, the interactions between matter and non-Hermitian light strongly modify the original NNN hopping parameter $t_2$. In particular, this hopping parameter splits into an anticlockwise hopping $t_2^\prime$ and a clockwise hopping parameter $t_2^{\prime \prime}$ that in the limit of $\gamma \neq 1$ are different (see illustration in panel (b) of Fig.~\ref{fig:6}). For the `anticlockwise' NNN hopping, $\phi_{ij}=\gamma \alpha$, where $\alpha=\cos(2\theta) \frac{g^2(1+g^2\gamma)}{(1+g^2)^2}\frac{V_0\pi}{\omega_c}$. From there, we can derive that:
\begin{align}
    &t_2^\prime=t_2 e^{\mathrm{i} \alpha\operatorname{Re}(\gamma) } e^{-\alpha \operatorname{Im}(\gamma)},\\
    &t_2^{\prime \prime}=t_2 e^{-\mathrm{i}  \alpha\operatorname{Re}(\gamma)} e^{\alpha\operatorname{Im}(\gamma) }.
\end{align}
This implies that the NNN hopping mechanism are now non-reciprocal and the hopping parameters acquire an imaginary part induced by the coupling to non-Hermitian chiral light.

Connecting to our original model described in Section~\ref{mod}, 
via a mapping defined as
\begin{align}
    &m=t_2 e^{-\alpha \operatorname{Im}(\gamma)} \cos[\alpha\operatorname{Re}(\gamma)]\\
    &a=t_2 e^{-\alpha \operatorname{Im}(\gamma)} \sin[\alpha\operatorname{Re}(\gamma)]\\
    &n=t_2 e^{\alpha \operatorname{Im}(\gamma)} \cos[\alpha\operatorname{Re}(\gamma)]\\
    &b=-t_2 e^{\alpha \operatorname{Im}(\gamma)} \sin[\alpha\operatorname{Re}(\gamma)],
\end{align}
we can directly match this system to the non-Hermitian model described in Section~\ref{mod}. More precisely, we derive that the chirality of light changes the sign of $\cos(2\theta)$, which in turn changes the sign of $a$ and $b$, thereby controlling the Chern number of the non-Hermitian honeycomb lattice. The non-Hermitian nature of the cavity makes the tight-binding Hamiltonian non-Hermitian and influences the sign of $m-n$. When the cavity exhibits dissipation, $\operatorname{Im}(\gamma)<0$, we naturally obtain $m>n$. The situation is reversed in the case of amplification.

In summary, we have demonstrated that by coupling an Hermitian tight-binding model in a honeycomb lattice to a non-Hermitian cavity we can obtain a non-Hermitian honeycomb lattice model and map it directly to the theoretical model considered in this manuscript. This suggests that our framework could, in principle, be realized in an experimental setup. By tuning the properties of the cavity, one could control both the topology and the non-Hermitian characteristics of the electronic spectrum.

\section{Discussion and Outlook}
In this work, we derive the semiclassical equations of motion that describe the dynamics of a wave packet with Gaussian shape, but also with more general wave-vector distribution, in non-hermitian topological systems characterized by complex Berry curvature. Our results show that the real part of the Berry curvature gives rise to an anomalous velocity, while the imaginary part induces an anomalous force -- features that are absent in Hermitian systems. We confirm the validity of the derived EOM by direct comparison with numerical simulations.

After a general analysis of the wave-packet dynamics, we focus on a non-Hermitian honeycomb topological lattice with nonreciprocal NNN hoppings, characterized by $m+\mathrm{i}a$ and $n+\mathrm{i}b$. We introduce the concept ``complex chirality" of NNN hoppings, defined as $S_\R={\rm sign}(n-m)$ and $S_\I={\rm sign}(b-a)$. When $a \neq b$, the sign of $S_\I$ dictates the Chern number of the bulk band gap and the handedness (counterclockwise/clockwise) of the edge states. In contrast, when $m \neq n$, the sign of $S_\R$ governs the imaginary Berry curvature, controlling the gain or loss of edge states, thus revealing a potential novel bulk-boundary correspondence.

Finally, we propose a potential experimental realization of the non-Hermitian honeycomb lattices using a non-Hermitian chiral cavity, which provides an intriguing platform to control the chirality and gain or loss of edge states. Additionally, we anticipate that future studies will focus on a comprehensive investigation of the precise mathematical proof for the novel bulk-boundary correspondence, as well as the development of a master equation approach to describe chiral light-matter coupling in dissipative cavities.

\appendix
\section{Quantum metric and fidelity number}
\label{app.A}
Let us consider an Hermitian system and define the distance between quantum states in wave vector space,
\begin{equation}
      ds^2_{\text{H}}=1-|\langle \Phi^i(\textbf{k})|\Phi^i(\textbf{k}+d\textbf{k}) \rangle |^2,
\end{equation}
where $|\Phi^i(\textbf{k})\rangle$ is the $i$th eigenstate of the Hermitian Hamiltonian.
For infinitesimal displacements $d\textbf{k}$, this distance can be written in terms of a quantum metric tensor (QMT) $g_{\mu \nu,\text{H}}^i$ (also called Fubini Study metric),
\begin{equation}
    ds^2_{\text{H}}=g_{\mu \nu,\text{H}}^i dk^\mu dk^\nu.\label{eq:distance}
\end{equation}
By ensuring gauge invariance, the QMT of the $i$th band can be defined as the real (symmetric) part of the quantum geometric tensor (QGT) \cite{Liu_2024,cheng2013},
\begin{equation}
\begin{aligned}
 Q_{\mu \nu,\text{H}}^i=&\left\langle\partial_\mu \Phi^i(\textbf{k}) \mid \partial_\nu \Phi^i(\textbf{k})\right\rangle\\&-\left\langle\partial_\mu \Phi^i(\textbf{k}) \mid \Phi^i(\textbf{k})\right\rangle\left\langle \Phi^i(\textbf{k}) \mid \partial_\nu \Phi^i(\textbf{k})\right\rangle.
\end{aligned} 
\end{equation}

Moving to a non-Hermitian system, left and right eigenvectors of a non-Hermitian Hamiltonian satisfy the biorthogonality condition $\langle \psi_L^i(\textbf{k}) | \psi_R^j(\textbf{k})\rangle=\delta_{ij}$ and the generalized completeness relations $\sum_i|\psi_R^i(\textbf{k})\rangle\langle \psi_L^i(\textbf{k}) |=\sum_i|\psi_L^i(\textbf{k})\rangle\langle \psi_R^i(\textbf{k}) | =I$~\cite{Brody_2014,PhysRevX.9.041015}. Since the bra and ket spaces are not Hermitian conjugates, the inner product must be redefined.

Therefore, the generalized quantum distance is now given by
\begin{align}
    ds^2&= (\langle \psi_L^i(\textbf{k}+d\textbf{k})|-\langle \psi_L^i(\textbf{k})|)(|\psi_R^i(\textbf{k}+d\textbf{k})\rangle - | \psi_R^i(\textbf{k})\rangle)\nonumber \\ &\approx \left\langle\partial_\mu \psi_L^i(\textbf{k}) \mid \partial_\nu \psi_R^i(\textbf{k})\right\rangle dk^\mu dk^\nu.
\end{align}
To ensure gauge invariance, the quantum geometric tensor (QGT) for a non-Hermitian system is finally defined as:
\begin{widetext}
\begin{align}
    Q_{\mu \nu}^i&=\left\langle\partial_\mu \psi_L^i(\textbf{k}) \mid \partial_\nu \psi_R^i(\textbf{k})\right\rangle-\left\langle\partial_\mu \psi_L^i(\textbf{k}) \mid \psi_R^i(\textbf{k})\right\rangle\left\langle \psi_L^i(\textbf{k}) \mid \partial_\nu \psi_R^i(\textbf{k})\right\rangle \nonumber \\&=
    \sum_j^{j \neq i} \frac{\left\langle \psi_L^i(\textbf{k})\left|\partial_{\mu} H(k)\right| \psi_R^j(\textbf{k})\right\rangle\left\langle \psi_L^j(\textbf{k})\left|\partial_{\nu} H(k)\right| \psi_R^i(\textbf{k})\right\rangle}{\left(E^i(\boldsymbol{k})-E^j(\boldsymbol{k})\right)^2}.
\end{align}
\end{widetext}

\begin{figure*}
\includegraphics[width=0.9\linewidth]{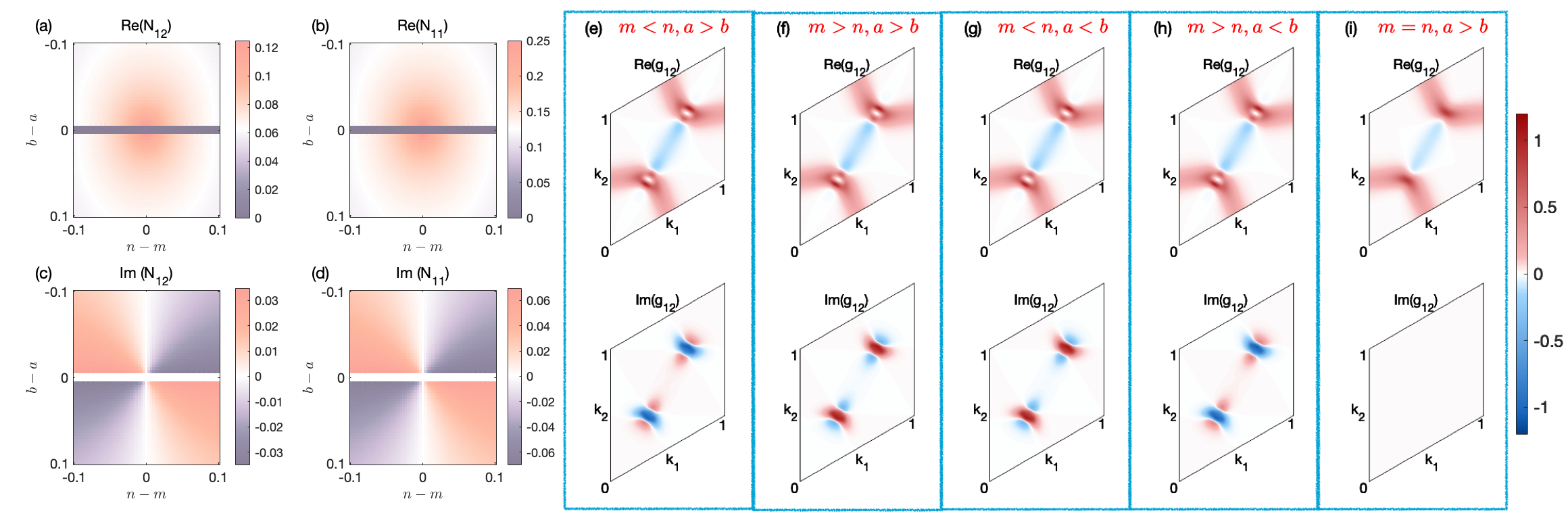}
    \caption{\textbf{Quantum metric tensor and fidelity number: }\textbf{(a-d)} Fidelity numbers, Eq. \eqref{fide}, for lower band as a function of the NNN hopping parameters. \textbf{(e-l)} Real and imaginary part of the lower-band quantum metric tensor $g_{12}$ for different hopping parameters: \textbf{(e)} $m$ = 0.1, $n$ = 0.2, $a$ = 0.2, $b$= 0.1; \textbf{(f)} $m$ = 0.2, $n$ = 0.1, $a$ = 0.2, $b$= 0.1; \textbf{(g)} $m$ = 0.1, $n$ = 0.2, $a$ = 0.1, $b$= 0.2; \textbf{(h)} $m$ = 0.2, $n$ = 0.1, $a$ = 0.1, $b$= 0.2; \textbf{(i)} $m$ = 0.1, $n$ = 0.1, $a$ = 0.25, $b$= 0.1;}
    \label{fig_appA}
\end{figure*}

 \begin{figure*}[ht]
\includegraphics[width=0.95\linewidth]{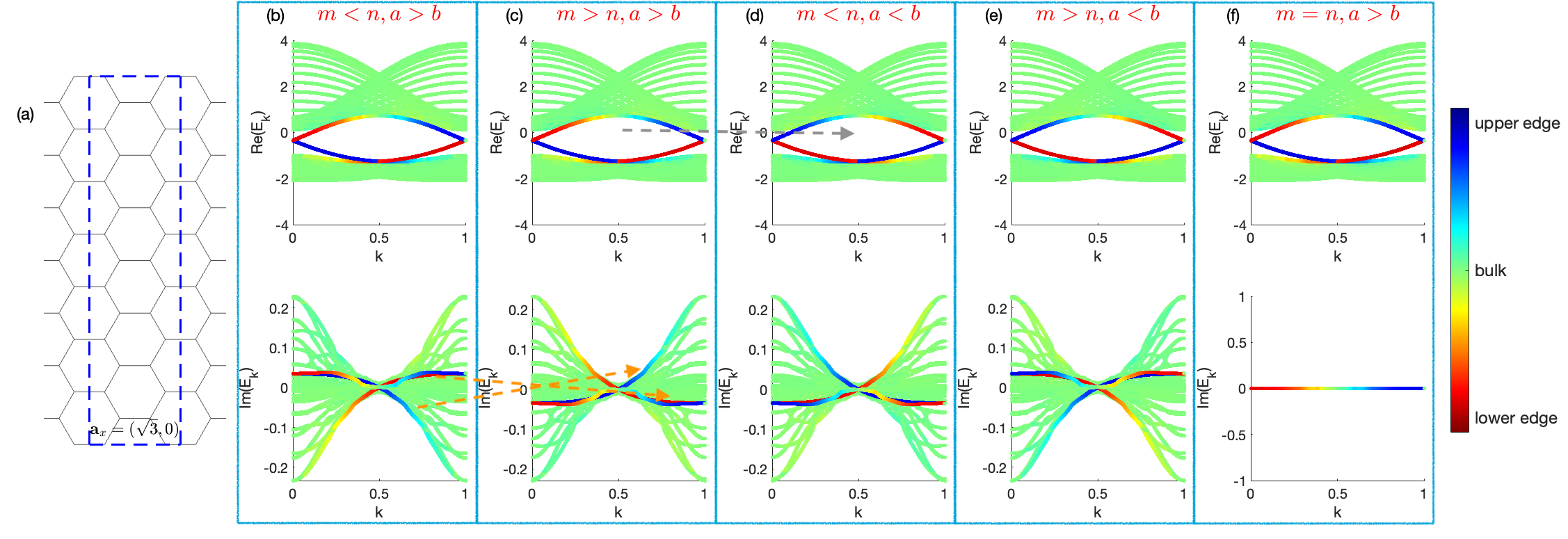}    \caption{\textbf{Bulk-edge correspondence and chiral edge states on armchair edges: }\textbf{(a)} An illustration of the honeycomb lattice strip with armchair edge. \textbf{(b-f)} Real and imaginary part of the energy spectrum for the strip with different hopping parameters: \textbf{(b)} $m$ = 0.1, $n$ = 0.2, $a$ = 0.1, $b$= -0.1; \textbf{(c)} $m$ = 0.2, $n$ = 0.1, $a$ = 0.1, $b$= -0.1; \textbf{(d)} $m$ = 0.1, $n$ = 0.2, $a$ = -0.1, $b$= 0.1; \textbf{(e)} $m$ = 0.2, $n$ = 0.1, $a$ = -0.1, $b$= 0.1; \textbf{(f)} $m$ = 0.15, $n$ = 0.15, $a$ = 0.1, $b$= -0.1; We set 40 atoms in one unit cell for the calculation. The blue/red energy bands represent eigenstates localized on the upper/lower edge of the strip, while the other states are bulk states. In all figures, we have set $a=-b$. The gray dashed arrow illustrates the inversion of the edge state chirality by changing the sign of $S_\I$. The orange dashed lines indicate the inversion of the $\operatorname{Im}( E_\textbf{k})$ of the chiral edge states with respect to the horizontal axes upon changing the sign of $S_\R$.}
    \label{fig_appB}
\end{figure*}

Using Schr\"{o}dinger's equation for right and left eigenvectors,
\begin{equation}
    H_\textbf{k} | \psi_R^i(\textbf{k})\rangle = \mathrm{i} \frac{\partial}{\partial t} | \psi_R^i(\textbf{k})\rangle,\,\, \langle \psi_L^i(\textbf{k})| H_\textbf{k} = -i \frac{\partial}{\partial t} \langle \psi_L^i(\textbf{k})|,
\end{equation}
we find that the time derivative of the QGT identically vanishes, $\partial Q/\partial t=0$. 
In general, the QGT is divided into a symmetric and an antisymmetric part, $Q_{\mu \nu}^i=g_{\mu \nu}^i-\frac{i}{2}B_{\mu \nu}^i$, where each part can be complex in non-Hermitian systems. The QMT, $g_{\mu \nu}^i=\frac{1}{2}(G_{\mu \nu}^i+G_{\nu \mu}^i)$, is the symmetric part of the QGT and it is real in Hermitian systems. This definition is also consistent with the expression for the generalized fidelity,
\begin{widetext}
\begin{align}
F^i(\textbf{k},\textbf{k}+d\textbf{k})&=\langle \psi_L^i(\textbf{k}) \mid \psi_R^i(\textbf{k}+d\textbf{k}) \rangle\langle \psi_L^i(\textbf{k}+d\textbf{k}) \mid \psi_R^i(\textbf{k}) \rangle\nonumber \\ &=1-g_{\mu \nu}^i d k^\mu d k^\nu,
\end{align}
\end{widetext}
provided in Ref. \cite{cheng2013}. 

Moreover, the Berry curvature is defined by the antisymmetric part of the QGT,
\begin{equation}
    B_{12}^i=(\nabla \times \textbf{A}^i)^z=\mathrm{i}(Q_{12}^i-Q_{21}^i),
\end{equation}
which is consistent with Eq.~\ref{eq:Berry}. For our system, $B_{12}^1=-B_{12}^2$.

We notice that previous studies
proposed different definitions of QGT, QMT, and Berry curvature in non-Hermitian systems  \cite{Hu:24,hu2024rolequantumgeometrictensor,PhysRevB.102.245147,PhysRevResearch.6.023202,wang2022anomalous}. Our definitions involve left and right eigenstates as bra and ket states, respectively. This choice ensures that $\partial Q/\partial t=0$.

Finally, the fidelity number is defined as the integral of the QMT
\begin{equation}
    N_{\mu \nu}^i=\int \frac{d^2 k}{(2 \pi)^2} g_{\mu \nu}^i(k),\label{fide}
\end{equation}
representing the average distance between quantum states. This quantity can also be interpreted as the overlap of Wannier states~\cite{chen1,chen2}. For our system, the fidelity number satisfies the relations $N_{1 2}^i=N_{2 1}^i$, $N_{1 1}^i=N_{2 2}^i$, and $N^1=N^2$ . 

We calculate the QMT and fidelity number of the lower band. Our results are shown in Fig.~\ref{fig_appA}. The dependence of the fidelity number on hopping parameters is depicted in panels (a-d) of Fig.~\ref{fig_appA}. Interestingly, the imaginary part of the QMT and the fidelity number changes sign with $S_\R$ and $S_\I$. We show the QMT in panels(e-i) of Fig.~\ref{fig_appA}.

\section{Chiral edge states on armchair edges}
\label{app.B}
We now consider a strip of the same honeycomb lattice with the armchair edge in the y-direction, as shown in panel (a) of Fig.~\ref{fig_appB}. The energy spectra are presented in panels (b-f) of Fig.~\ref{fig_appB}.

The chirality of the edge states on armchair edges, determined by the Chern number, is consistent with that of the edge states on zigzag edges, as shown in panels (b-f) of Fig.~\ref{fig:4} and Fig.~\ref{fig_appB}. More precisely, the chirality is not influenced by the shape of the edge. When $S_\I<0$/$S_\I>0$, anticlockwise/clockwise propagating topological edge states exist.

Similarly to the spectrum for zigzag edges, $S_\R$ controls the behavior of $\operatorname{Im}(E_k)$. In Fig.~\ref{fig_appB}, we set $a=-b$ so that the Hamiltonian is Hermitian and $\operatorname{Im}(E_k)=0$ when $m=n$. Then, the shape of $\operatorname{Im}(E_k)$ flips relative to the horizontal axes, $\operatorname{Im}(E_k)$=0,  when $S_\R$ changes sign. When $a\neq -b$, $\operatorname{Im}(E_k)$ does not simply change sign with $S_\R$, but $S_\R$ still influences the behavior of $\operatorname{Im}(E_k)$.

\section{Non-Hermitian chiral cavity}
\label{app.C}

By using the time dependent vector potential $\hat{\mathbf{A}}(t)=A_0\left(\boldsymbol{f} \hat{a} e^{-\mathrm{i} \omega_c t}+\gamma\boldsymbol{f}^* \hat{a}^{\dagger}e^{\mathrm{i} \omega_c t}\right)$, we obtain the electric field as
\begin{equation}
    \hat{\mathbf{E}}(t)=-\frac{\partial \hat{\mathbf{A}}(t)}{\partial t}=\mathrm{i} \omega_c A_0\left(\boldsymbol{f} \hat{a} e^{-\mathrm{i} \omega_c t}-\gamma\boldsymbol{f}^* \hat{a}^{\dagger}e^{\mathrm{i} \omega_c t}\right).
\end{equation}
Since only $\hat{a}^\dagger \hat{a}$ and $\hat{a} \hat{a}^\dagger$ give finite contributions to $\langle n| \hat{\mathbf{E}}^2|n\rangle$, the energy associated to the electric field is given by
\begin{equation}
    \begin{aligned}
        \frac{\epsilon_0}{2} \int_V dV \langle  n|\hat{\mathbf{E}}^2 |n \rangle &= \gamma \frac{\epsilon_0}{2} V  \omega_c^2  A_0^2\langle n|  ( \hat{a}^{\dagger} \hat{a} + \hat{a}\hat{a}^{\dagger}) | n \rangle\\
        &=\gamma \epsilon_0 V  \omega_c^2  A_0^2 (n+\frac{1}{2}),
    \end{aligned}
\end{equation}
where $n$ is the quantum number of the photonic mode follows $\hat{a}^{\dagger} \hat{a} |n \rangle=n|n \rangle$.
The energy of the electric/magnetic field  is then given by
\begin{equation}
    E=2\gamma \epsilon_0 V  \omega_c^2  A_0^2(n+\frac{1}{2})=\hbar \omega_c (n+\frac{1}{2}).
\end{equation}
Thus, $A_0=\sqrt{\frac{\hbar}{2\omega_c V  \epsilon_0 \gamma}}$. We can set $\omega_c=\frac{|\omega_c|}{\gamma}$ and $|\gamma|=1$ to make $A_0$ real. Thus $\omega_c$ is complex when $\gamma$ is complex. Notice that sign($\operatorname{Im}(\gamma)$) controls the topology and non-Hermiticity of the tight-binding honeycomb lattice, as explained in Section~\ref{sec:realization}. Here, positive/negative $\operatorname{Im}(\gamma)$ corresponds to negative/positive $\operatorname{Im}(\omega_c)$, which relates to the dissipation/amplification of the cavity, respectively. We choose $\operatorname{Im}(\gamma)>0$ to retain the stability of the system.

We then apply a similarity transformation $\hat{U}=\exp \left[-\mathrm{i} \frac{\xi}{\hbar} \hat{\mathbf{p}} \cdot \hat{\boldsymbol{\pi}}\right]$ to the Hamiltonian in Eq.~\ref{eq:H}:
\begin{widetext}
\begin{equation}
    \begin{aligned}
\hat{H}_U&=\hat{U}^{-1} \hat{H} \hat{U} =\frac{\hat{\mathbf{p}}^2}{2 m_{\mathrm{eff}}}+V\left(\mathbf{r}+\xi \hat{\boldsymbol{\pi}}-\gamma\cos(2\theta)\frac{\xi^2}{2 \hbar} \hat{\mathbf{p}} \times \mathbf{e}_{\mathbf{z}}\right)+\hbar \omega_{\mathrm{eff}} \hat{a}^{\dagger} \hat{a},\\
m_{\mathrm{eff}}&=m\left(1+g^2\gamma\right),\quad 
\omega_{\mathrm{eff}}=\omega_c\left(1+g^2\right),
    \end{aligned}
\end{equation}
\end{widetext}
where $\xi=\frac{qA_0}{m \omega_{\mathrm{eff}}}=\frac{g}{1+g^2} \sqrt{\frac{\hbar}{m \omega_c}}$ to eliminate the linear term $\hat{\mathbf{p}} \cdot \hat{\mathbf{A}}$, and $g$ is set to be $g=\sqrt{\frac{\left(q A_0\right)^2}{m \hbar \omega_c}}$, which quantifies the strength of the light-matter coupling. We use the photonic momentum operator $\hat{\bm{\pi}}=\mathrm{i}\left(\gamma\boldsymbol{f}^* \hat{a}^{\dagger}-\boldsymbol{f} \hat{a}\right)$ and its commutation relation $\left[\hat{\bm{\pi}}_x, \hat{\bm{\pi}}_y\right]=i \gamma\cos(2\theta)$ in the derivation. Notice that $\hat{U}^{-1}=\exp \left[\mathrm{i} \frac{\xi}{\hbar} \hat{\mathbf{p}} \cdot \hat{\boldsymbol{\pi}}\right]$, and does not coincide with $\hat{U}^{\dagger}$ in non-Hermitian systems. Moreover, we also make use of the Hadamard's lemma, $e^{\hat{A}} \hat{B} e^{-\hat{A}}=\hat{B}+[\hat{A}, \hat{B}]+\frac{1}{2!}[\hat{A},[\hat{A}, \hat{B}]]+\ldots$, in the derivation.

Finally, we expand the potential $V$ into the first-order in $\xi$ to obtain
\begin{equation}
\begin{aligned}
   \hat{H}_U & \approx \frac{(\hat{\mathbf{p}}-q \mathbf{A(\mathbf{r})})^2}{2 m_{\mathrm{eff}}}+V\left(\mathbf{r}\right)+\hbar \omega_{\mathrm{eff}} \hat{a}^{\dagger} \hat{a}+\nabla V(\mathbf{r}) \xi \hat{\pi},\\
    \mathbf{A(\mathbf{r})}  & = \gamma\cos(2\theta) ~\frac{m_{\mathrm{eff}} \xi^2}{2 q \hbar} (\mathbf{e}_{\mathbf{z}} \times \nabla V(\mathbf{r})).
\end{aligned}
\end{equation}
Following all these steps, one can recover the result presented in the main text, Eq. \eqref{pp}.

\end{document}